# Using synchrotron X-ray scattering to study the diffusion of water in a weakly-hydrated clay sample

Y. Méheust[a]*, B. Sandnes[b], G. Løvoll[b], K. J. Måløy[b], J. O. Fossum[a],
G. J. da Silva[c], M. S. P. Mundim[c], R. Droppa[d], D. d. Miranda Fonseca[a]

[a] *Department of Physics, Norwegian University of Science and Technology, Hoegskoleringen 5, NO-7491 Trondheim, Norway*

[b] *Physics Department, University of Oslo, Postboks 1048 Blindern, NO-0316 Oslo, Norway*

[c] *Physics Department, University of Brasilia, Caixa Postal 04455, 70919-970 Brasília – DF, Brazil*

[d] *The Brazilian Synchrotron Laboratory (LNLS), Caixa Postal 6192, CEP 13084-971 Campinas – SP, Brazil*

## ABSTRACT

We study the diffusion of water in weakly-hydrated samples of the smectite clay Na-fluorohectorite. The quasi one-dimensional samples are dry compounds of nano-layered particles consisting of ~ 80 silicate platelets. Water diffuses into a sample through the mesoporosity in between the particles, and can subsequently intercalate into the adjacent particles. The samples are placed under controlled temperature. They are initially under low humidity conditions, with all particles in a 1WL intercalation state. We then impose a high humidity at one sample end, triggering water penetration along the sample length. We monitor the progression of the humidity front by monitoring the intercalation state of the particles in space and time. This is done by determining the characteristic spacing of the nano-layered particles in situ, from synchrotron wide-angle X-ray scattering measurements. The spatial width of the intercalation front is observed to be smaller than 2mm, while its velocity decreases with time, as expected from a diffusion process.

## INTRODUCTION

Fluorohectorite is a synthetic smectite with formula $X_x(Mg_{3-x}Li_x)Si_4O_{10}F_2$ per half unit cell, where X is a cation (Na, Ni, Li, Fe). It is polydisperse, with platelets sizes ranging from a few tenth of nm up to a few micrometers, and has a large surface charge[1] (1.2 e- per unit cell, against 0.6 for montmorillonite) which causes platelets to remain stacked in water suspensions, even in low saline environment. These stacks are strengthened by the presence of the intercalated cation X, which is shared by two adjacent silica sheets in the stack. They contain on average around 80 platelets[2]. As for natural smectites, water molecules can also intercalate inside theses nano-layered particles, causing them to swell in a stepwise molecular packing process[3-5] where some configurations are thermodynamically favoured and interpreted as "water layers" successively intercalated in the stacks.

E-mail of the corresponding author: meheust@phys.ntnu.no

For Na-fluorohectorite (X=Na), which we study here, we observed no more than 3 intercalated water layers[6]. The hydration state of a nanolayered particle depends on the temperature and surrounding humidity. Characteristic platelet separation[6] (d=1.0, 1.2 and 1.5 nm) and transition temperatures[7] under conditions of low (~5 %) and high (~ 98 %) relative humidity have been determined for the three hydration states, respectively.

In this paper we study the macroscopic diffusion of water into weakly hydrated Na-fluorohectorite samples. These samples, obtained by dehydration of water suspensions, are dry assemblies of the nano-layered particles described above. Water diffuses inside of them through the mesoporosity in between the particles. Applying a gradient of vapor partial pressure between the two ends of quasi-dimensional samples, we study this water diffusion, at the macroscopic scale. Depending on the temperature, a particle inside the assembly is likely to swell as the water diffusion front reaches it, due to the increase in the humidity level of the surrounding mesoporous space. Under those conditions, the diffusion of water through the mesoporous space goes along with the progression of an intercalation (or swelling) front inside the sample. We monitor the progression of this intercalation front by repeatedly recording the characteristic platelet separation, d, at distances regularly spaced from the wet sample end. The value for d is obtained from in situ wide angle X-ray scattering measurements.

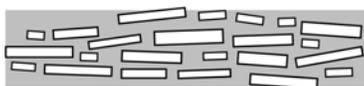

Fig. 1: 2D view of the sedimented samples' geometry. The mesoporous space is denoted by the gray shading. The clay particles have their silica sheets oriented along the horizontal direction, on average. The RMS deviation for the alignment of individual particles from this mean orientation is between 15 and 16.5° [11].

Many studies of water transport in a porous medium have used NMR[8,9]. This is to our knowledge the first study of such a system using wide angle X-ray scattering.

## SAMPLES

The samples were prepared in the following way. Raw fluorohectorite powder was purchased from Corning Inc. (New York). It was dissolved in deionized water; the suspension was stirred for several days. NaCl was then added in an amount ~ 10 times larger than the estimated amount of interlayer charges in the suspensions, so as to force the replacement of all intercalated cations by $Na^+$. This caused the suspension to flocculate. After 2 weeks of stirring, it was left to sediment, and the supernatant was removed. Excess ions were subsequently removed by placing the flocculated clay in dialysis membranes and in contact with deionized water, which was changed every second day, until a test using $AgNO_3$ showed not presence of remaining $Cl^-$ ions. At this point, two types of samples were prepared.

(i) Part of the sediment was heated for 6 hours at 120°C. The resultant powder was grinded in a mortar. The grinded powder was then placed in X-ray glass capillaries, with a diameter 2mm and a wall thick 0.01mm. They were vibrated while filling them up so as that the compaction of the powder inside the tube be as uniform as possible. We refer to these samples as "powder samples".

(ii) Part of the sediment was suspended in deionized water again, and after stirring, water was expelled by heating up. The resulting assemblies were then cut in strips of 4mm by 47mm. We refer to these samples as "sedimented samples". In contrast to the powder samples that are isotropic, these samples are anisotropic, with a marked average alignment of the clay particles parallel to the horizontal plane[10,11] (see Fig. 1).

## EXPERIMENTAL METHOD

The experiments consist in repetitive in situ measurements of the proportions of particles in the 1WL- and 2WL-hydration states, at positions regularly spaced along the sample length. This is done by positioning these points of the sample in front of a horizontal X-ray beam, and recording one-dimensional scattering spectra for each position and time.

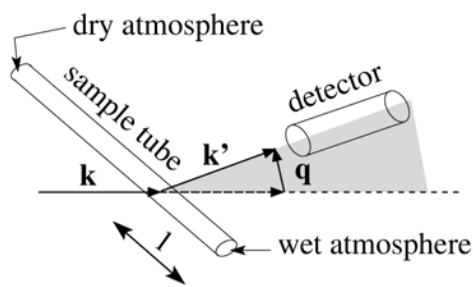

Fig. 2: Sketch of the scattering geometry, featuring the momentum of the incident photons, **k**, that of the scattered photons, **k'**, and the momentum transfer **q**. These 3 vectors are in a vertical plane (denoted by the gray shadings) in which the detectors moves to record the scattering spectra. The sample can be translated along its length so as to vary the distance l from the wet end at which the beam hits the sample.

*X-ray scattering setup*

Wide angle scattering (WAXS) experiments were carried out at the D12A XRD1 beamline at LNLS (the Brazilian Synchrotron Laboratory). The Si(111) two-crystal monochromator with horizontally focusing, coupled to a vertically focusing Rh-coated X-ray mirror, provided a 1x1mm² or 1x3 mm² beam with an energy E = 10.4 kV and a dispersion on the energy $\Delta E=E/1170$. The sample was placed horizontally, with its length perpendicular to the incoming X-ray beam, and a scattering spectrum was recorded in a vertical plane, using a 3-circle Huber diffractometer and a NaI scintillation point detector. The scattering geometry is shown in Fig. 2.

*Sample holder*

The sample was placed on a thermally-controlled copper block, as shown in Fig. 3(a), with heat-conducting paste in between them. The temperature control setup allowed both cooling and heating of the copper block, and consisted of a Peltier element placed under the block, a thermistor to measure its temperature, and a computer-controlled PID system to adjust the Peltier's excitation to the measured temperature. The Peltier was applied a reference temperature on its side opposite to the copper block, in the form of water circulating from a regulated water bath. The temperature of the bath could be changed so as to increase the range of temperature available in the copper block. The overall precision of the temperature control system was found to be around 0.01 K.

The sample environment was also controlled in humidity. It was sealed along its length and in contact with two reservoirs at its ends. Air with a controlled humidity was circulated separately in the two reservoirs, which allowed to impose different controlled humidity levels at the sample ends, and therefore to impose a humidity gradient inside it. A picture of the sample holder as it was during the experiment is presented in Fig. 3(b). It features the sample (i) on its holder, the two pairs of plastic tubing (ii) at each end of the sample, and the temperature sensor (iii).

*Experimental protocol*

The experiments were initiated in the following way. Prior to the scattering measurements, the samples were maintained at a temperature of 24°C, with dry air circulating at both ends, long enough for all scatterers to be at equilibrium under a 1WL hydration state. Reference scattering spectra were then recorded. The spectrum labeled "1WL" in Fig. 4 is an example of such a graph. Temperature was then lowered down to 5°C. Half an hour later, thermal equilibrium was reached and we started circulating humid air at one end of the sample, imposing a humidity gradient across the length of the sample. This was defined as the initial time for water diffusion. Water penetrating the sample came in contact with particles in the 1WL state at a temperature where their equilibrium state is 2WL, hence triggering the displacement of an intercalation front in the sample, along with water diffusion.

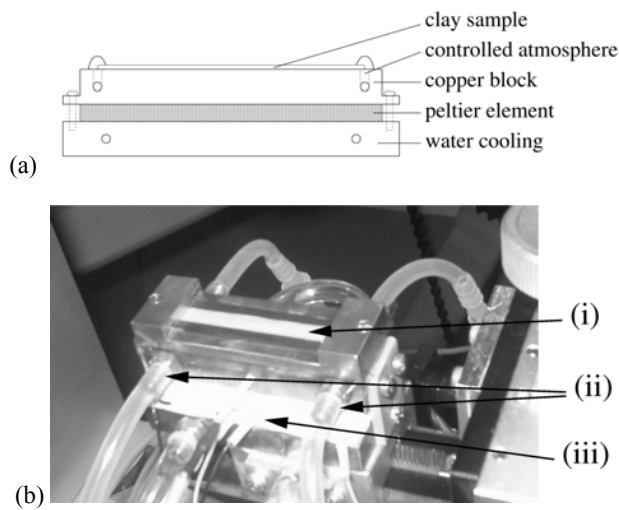

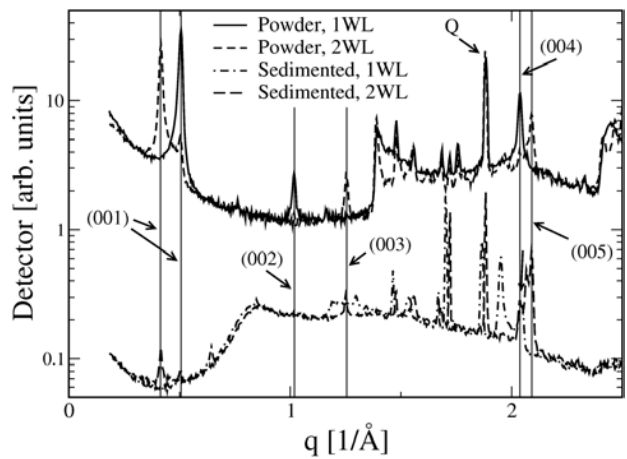

Fig. 3: (a) Side view of the sample holder, with copper block and Peltier element underneath - (b) Photo of the sample environment during an experiment carried out on a sedimented sample. The sedimented sample is isolated from the lab atmosphere by a thin kapton film glued on top of it.

Fig. 4: Scattering spectra as a function of the momentum transfer q for powder (spectra at the top) and sedimented samples (spectra at the bottom). In each case, two spectra are shown: one recorded at a position in the sample where most of the scatterers are in the 1WL hydration state, the other at a position where most are in the 2WL state. The vertical scale is logarithmic, the spectra have been normalized arbitrarily for conveniency.

Pressure was measured in the chambers at both ends of the samples prior to- and at the end of- the experiment. No significant pressure gradient was measured between the sample ends.

## X-RAY DATA

*Scattering from the nano-layered particles*

The one-dimensional scattering spectra display sharp peaks for deviation angles 2θ corresponding to a Bragg reflection of some of the scatterers in the scattering volume. Apart from peaks from quartz impurities (denoted by "Q" in Fig. 4), these sharp peaks are mostly those characteristic of Bragg planes associated with the particle stacks. In Fig. 4, we show reference spectra for scattering volumes where the scatterers are in the 1WL and 2WL hydration states, and for the two types of samples. The various (00k) orders observed are indicated. Due to form factor effects, some of those orders are extinct. The spectra for the sedimented samples exhibited a large background caused by the silicon glue that we used to seal the top of the samples to the kapton. This was not observed for the powder samples, as they were inside capillary tubes.

The (001) peaks are very weak for the sedimented samples, this is due to some shadowing effect by the copper block at low angle as these samples are actually lying in a trench carved onto the top surface of the copper block. For that reason we had to treat higher order peaks when working with the sedimented samples, while for powder samples we could use the first order peaks, which are also the better resolved ones. For that reason we only present, in what follows, data obtained from the powder samples.

*Hydration transition at a given position*

Fig. 5(a) is a close view of the spectrum measured at a position l = 7.0 mm from the wet end of the sample, at different times after the diffusion process has started. The width of the beam was 3 mm. As water diffuses inside the sample, the relative humidity in the scattering volume increases. Consequently, water starts intercalating in some of the scattering particles. This is marked by a decrease in the intensity of the peak characteristic for the pure 1WL hydration state, and the appearance of an asymmetry in its shape. This asymmetry evolves into a broad and low intensity peak that appears between the 1WL and 2WL peaks. This is characteristic of the existence of scatterers in various coexisting mixed Hendricks-Teller intercalation states, with different proportions of 1WL and 2WL spacings inside scatterers[12]. These mixed states progressively blend into the shoulder of the pure 2WL peak that has started appearing after 7 hours. After ~ 17 h, only a pure 2WL peak is observed. We therefore consider that the intercalation front has penetrated in the scattering volume at t = 7.0 h, and left it 10±1 hours later.

From Fig. 5(a), we have plotted the evolution of the relative intensities for the pure 1WL and 2WL peaks as a function of time (see Fig. 5(b)). This was done by first subtracting the background from Fig. 5(a), and then normalizing the amplitudes by those observed when no random intercalation is present. From the intersection of the two curves, we infer that the time at which the mean front position is in the middle of the scattering volume is ~ 8.7±1 h, at l = 7±1 mm.

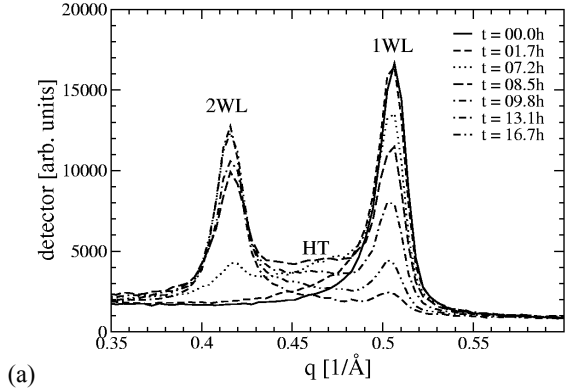

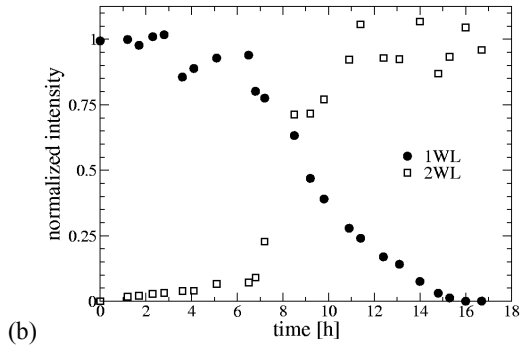

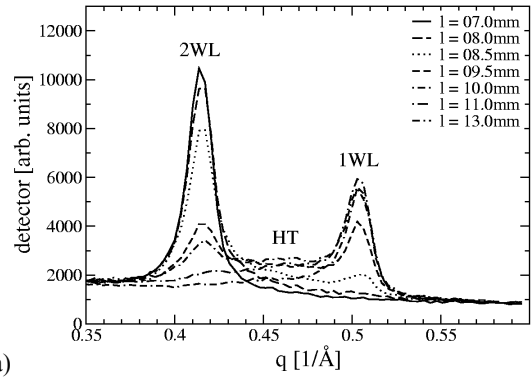

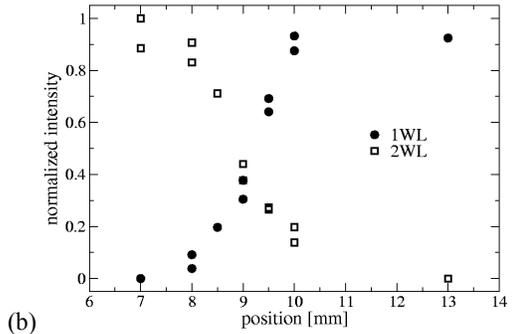

Fig. 5: (a) Close view of the scattering spectra recorded at l = 7.0 mm from the wet end of the sample, at various times between 0 and 16.7 h. As the intercalation travels through the scattering volume, the spectrum continuously moves from a pure 1WL peak to a pure 2WL peak. - (b) Normalized relative intensities of the pure 1WL and 2WL peaks in (a), as a function of time.

Fig. 6: (a) Close view of the scattering spectra recorded at time t = 19 h from the beginning of the water diffusion, and at several distances l from the wet end of the sample. - (b) Normalized relative intensities of the pure 1WL and 2WL peaks in (a), as a function of l. A transition from 2WL to 1WL is observed for increasing values of l.

*Hydration transition at a given time*

Fig. 6(a) shows the evolution of the 001 peaks at a given time, t = 19h, and for various positions l from the wet sample end. When traveling from the wet to the dry end of the sample, a similar transition is observed from the pure 2WL state to the pure 1WL state, through Hendricks-Teller states of random intercalation. In Fig. 6(b), we have plotted the relative intensities of the pure peaks in Fig. 6(a), as a function of the distance l. This provides us with a "spatial picture" of the intercalation front. However, the apparent width of the front in Fig. 6(b), around 3mm, results from the convolution of the front profile by the width of the X-ray beam, which is this case was 1mm. Hence, we can conclude that the intercalation front is quite sharp: less than 2mm in width. We also observe that the mean spatial position of the front at t = 19±1 h is l = 9.1±0.5 mm.

## DISCUSSION

In each of the two previous sections, we have obtained one estimate of the intercalation front position as a function of time: 7.0±1 mm after 8.7±1 h, and 9.1±0.5 mm after 19.0±1 h. We can use the estimate of the front width (2mm) obtained in the previous section to obtain two more estimates. Supposing that the front penetrates the scattering volume at the time when the 2WL peak in Fig. 5(a) starts appearing, and taking the beam width into account, we infer a mean position of the front of 4.5±1mm after 7.2±1 h. In the same way we can assume that when the front leaves the scattering volume after 17±1 h, his mean position if 9.5±1 mm. Those 2 points, plus the initial point (0h,0mm), are plotted in Fig. 7.

How the progression of a sharp intercalation front is related to the penetration of water molecules inside the mesoporosity remains an open question. A possible mechanism would be that the vapor partial pressure in the scattering volume need to reach a certain triggering value before significant intercalation is observed. The intercalation front would then occur in the vicinity of the iso-humidity line defined by that particular value of the vapor partial pressure. Since there is no pressure difference between the two sample ends, the driving force of the water transport is the gradient of relative humidity, or water partial pressure, along the sample length. The water transport is thus expected to be a diffusive process. Let us consider a standard diffusive process with boundary conditions corresponding to our experimental configuration, that is: (i) a constant humidity difference between the two sample ends, and (ii) an initial configuration where the whole sample is at the lowest humidity. If the sample length is infinite, the solution is well known: the humidity profile has as shape defined by an error function, and its position varies as a square root function of t; so does the position of any iso-humidity line. In our experiment, due to the finite sample length, the humidity profile is expected to evolve from a step function at t = 0 to a linear profile between the two boundary values at infinite time.

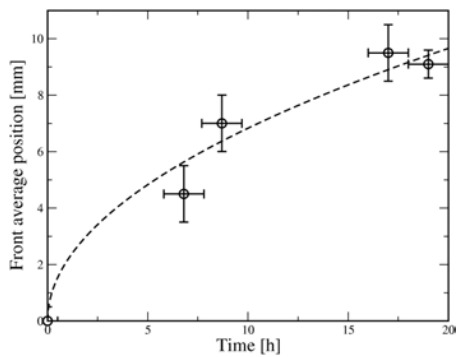

Fig. 7: Position of the mean intercalation front position as a function of time. The dashed line is a fit to the behavior expected from a standard diffusion process, where $l \sim t^{1/2}$.

Consequently, if intercalation is triggered by a threshold humidity value $h_c$, the intercalation front is expected to stop at the position $l_c$ where the final humidity profile reaches $h_c$. Hence, the evolution of the front position as a function of time does not follow a square root law. However, at initial times, as the dry end of the sample is not "felt" by the water molecules entering at the wet end, this evolution should be close to a square root law. For the data plotted in Fig. 7, the largest probed l value is smaller than the 4$^{th}$ of the overall sample length, hence we believe that the standard diffusive transport through an infinite sample could be a first approximation description. The dashed line in Fig. 7 corresponds to a square root law that best fits the data; in the limit of the large uncertainties, it is consistent with the data.

Let us point out that in our experiment, water molecules that enter the sample do not remain in the mesoporosity, but intercalate into clay particles, which act as a sink for diffusing water. In addition, the subsequent swelling of clay particles modifies the configuration of the porous space, diminishing its porosity. Those two factors should have an impact on the transport, so diffusion process is expected to be slightly different from what is described by a standard diffusion equation. Obviously, uncertainties and data scarcity do not allow addressing that discrepancy in the current study.

## CONCLUSION AND PROSPECTS

We have studied the diffusion of water inside quasi one-dimensional weakly-hydrated samples of Na-fluorohectorite, after imposing different humidity levels at their two ends. We used WAXS measurements to monitor the displacement of an intercalation front as water diffuses in the mesoporosity of the samples. The front was observed to be less than 2mm wide when it reached a distance l = 9mm from the wet side of the sample. In the range of l values investigated, its progression as a function of time is in first approximation consistent with a diffusive law for an infinite sample: $t^{1/2}$.

We shall carry out more extended measurements of the intercalation front position as a function of time, in order to get a better description of its displacement in time. The use of a micro-focus beam, allowing the removal of the uncertainty due to the beam size, is planned.

In order to better understand the connection between the intercalation front and the diffusion of water in the mesoporous space, we plan to precisely determine which relative humidity value (or vapor partial pressure value) triggers the intercalation of a second layer of water. This shall be done through carefully temperature- and humidity-controlled intercalation experiments on a very small amount of clay powder, with gravimetry and wide angle X-ray scattering measurements. With those measurements coupled to microfocus WAXS diffusion experiments, we eventually hope to be able to confront our data to theories for anomalous scattering[13-14], like in a recent study on related systems[15].

The results presented here were obtained on powder samples. For reasons explained earlier, sedimented samples, where clay particles exhibit a marked alignment along the horizontal plane, did not provide data where the 001 peaks could be analyzed. Another prospect of the study is to solve this technical problem in order to compare the typical diffusion time scale for those samples, as opposed to the isotropic samples.

## ACKNOWLEDGEMENTS

We are grateful to the staff at LNLS for their technical support during the WAXS experiments. We also acknowledge helpful discussions with Kenneth D. Knudsen. This work is supported by the Research Council of Norway (RCN) through funding granted in the framework of the RCN Nanomat Program and the RCN Strategic University Program.